\documentclass[sigconf]{acmart}

\AtBeginDocument{%
  \providecommand\BibTeX{{%
    \normalfont B\kern-0.5em{\scshape i\kern-0.25em b}\kern-0.8em\TeX}}}

\setcopyright{acmlicensed}
\copyrightyear{2024}
\acmYear{2024}
\acmDOI{XXXXXXX.XXXXXXX}

\acmConference[MM'24]{}{October 28 - November 1, 2024}{Melbourne, Australia.}

\acmISBN{978-1-4503-XXXX-X/18/06}
\usepackage{array}
\usepackage{url}
\usepackage{multirow}
\usepackage{balance}
\usepackage{bbm}
\usepackage{bbding}
\usepackage{xcolor}
\usepackage{caption}
\usepackage{subfigure}
\usepackage{setspace}
\usepackage{multirow}
\usepackage[T1]{fontenc}
\usepackage{enumerate}
 
\begin{document}

\title{Multi-Stage Face-Voice Association Learning with Keynote Speaker Diarization}

\author{Ruijie Tao}
\affiliation{%
  \institution{National University of Singapore}
  % \streetaddress{1 Th{\o}rv{\"a}ld Circle}
  % \city{Singapore}
  % \country{Singapore}
  \city{}
  \country{}
  }
\email{ruijie@nus.edu.sg}

\author{Zhan Shi}
\affiliation{%
  \institution{The Chinese University of Hong Kong}
  % \streetaddress{1 Th{\o}rv{\"a}ld Circle}
  % \city{Shenzhen}
  % \country{China}
  \city{}
  \country{}
  }
\email{zhanshi1@link.cuhk.edu.cn}

\author{Yidi Jiang} 
\authornote{Corresponding author.}
\affiliation{%
  \institution{National University of Singapore}
  % \streetaddress{1 Th{\o}rv{\"a}ld Circle}
  % \city{Singapore}
  % \country{Singapore}
  \city{}
  \country{}
  }
\email{yidi_jiang@u.nus.edu}

\author{Duc-Tuan Truong}
\affiliation{%
  \institution{Nanyang Technological University}
  % \streetaddress{1 Th{\o}rv{\"a}ld Circle}
  % \city{Singapore}
  % \country{Singapore}
  \city{}
  \country{}
  }
\email{truongdu001@ntu.edu.sg}

\author{Eng-Siong Chng}
\affiliation{%
  \institution{Nanyang Technological University}
  % \streetaddress{1 Th{\o}rv{\"a}ld Circle}
  % \city{Singapore}
  % \country{Singapore}
  \city{}
  \country{}
  }
\email{aseschng@ntu.edu.sg}

\author{Massimo Alioto}
\affiliation{%
  \institution{National University of Singapore}
  % \streetaddress{1 Th{\o}rv{\"a}ld Circle}
  % \city{Singapore}
  % \country{Singapore}
  \city{}
  \country{}
  }
\email{massimo.alioto@nus.edu.sg}

\author{Haizhou Li}
\affiliation{%
  \institution{The Chinese University of Hong Kong}
  % \streetaddress{1 Th{\o}rv{\"a}ld Circle}
  % \city{Shenzhen}
  % \country{China}
  \city{}
  \country{}
  }
\email{haizhouli@cuhk.edu.cn}

\renewcommand{\shortauthors}{Ruijie Tao, Zhan Shi and Yidi Jiang, et al.}
%\name{Ruijie Tao$^1$, Shi Zhan$^2$, Yidi Jiang$^1$, Duc-Tuan Truong$^3$, Eng-Siong Chng$^3$, Massimo Alioto$^1$ and Haizhou Li$^{1,2,4}$}

%\address{\quad$^1$National University of Singapore, Singapore\quad$^2$The Chinese University of Hong Kong, Shenzhen, China \\ $^3$Nanyang Technological University, Singapore\quad$^4$ Shenzhen Research Institute of Big Data, Shenzhen, China}

\begin{abstract}
The human brain has the capability to associate the unknown person's voice and face by leveraging their general relationship, referred to as ``cross-modal speaker verification''. This task poses significant challenges due to the complex relationship between the modalities. In this paper, we propose a ``Multi-stage Face-voice Association Learning with Keynote Speaker Diarization''~(MFV-KSD) framework. MFV-KSD contains a keynote speaker diarization front-end to effectively address the noisy speech inputs issue. To balance and enhance the intra-modal feature learning and inter-modal correlation understanding, MFV-KSD utilizes a novel three-stage training strategy. Our experimental results demonstrated robust performance, achieving the first rank in the 2024 Face-voice Association in Multilingual Environments (FAME) challenge with an overall Equal Error Rate (EER) of 19.9\%.
Details can be found in \textcolor{magenta}{\url{https://github.com/TaoRuijie/MFV-KSD}}.

\end{abstract}

%%
%% The code below is generated by the tool at http://dl.acm.org/ccs.cfm.
%% Please copy and paste the code instead of the example below.



\maketitle

\section{Introduction}
\begin{figure}[!ht]
    \centering
    \includegraphics[width=\linewidth]{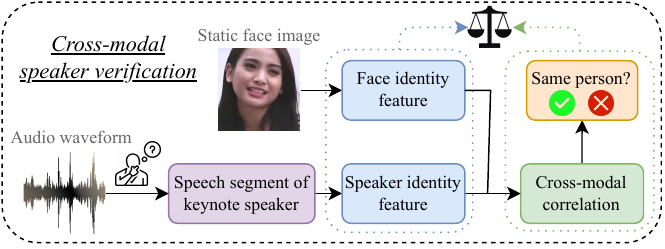}
    \caption{Cross-modal speaker verification system requires to focus on the active speech segment of ``keynote speaker'' (who talks most), and keep a balance between single-modal understanding and cross-modal correlation learning.}
    \label{fig:cover}
\end{figure}
Speaker recognition aims to verify the identity of a speaker based on their unique voice characteristics. Deep learning-based speaker representation systems, with the assistance of large-scale datasets, have achieved robust performance~\cite{ssl_app_sv, ssl_app_age_1, liu2024neural,ssl_app_add}. Recent studies have demonstrated that this verification task can also be extended to the cross-modal scenarios~\cite{cross_modal_is2020}. Specifically, even for the unknown identity, the human brain can recognize whether a static facial image and a heard voice belong to the same individual, relying on general cross-modal relationships such as gender, nationality, and age~\cite{oh2019speech2face,chung2020perfect,nagrani2018seeing}. Various related studies have been explored for audio-visual speech processing. For instance, FaceFilter~\cite{chung2020facefilter} studied audio-visual speech separation conditioned on a still face image of a target speaker. Face-based target speech diarization~\cite{jiang2024target} employed a static face image as the prompt to specify the desired speaker.

This year, the FAME Challenge \cite{saeed2024face} focuses on enhancing the integration of face and voice embeddings for robust speaker recognition in multilingual contexts. Instead of relying solely on cross-validation within a single language, FAME evaluates the effects of multiple languages on the verification process. 

Although the existing cross-modal solutions have achieved remarkable improvement by leveraging the deep learning methods, there are still two main problems in this task: As shown in Figure~\ref{fig:cover}, from the data aspect, in real-world scenarios, raw audios often contain conversational or non-speech segments. However, speaker recognition models exhibit robust capabilities in capturing features from single-modal inputs~\cite{liu2023disentangling,tao2024voice,chen2022large}. The noisy inputs will significantly degrade the ability of speaker models to extract accurate embeddings. Meanwhile, it is challenging to find a suitable solution: traditional speaker diarization methods can hardly distinguish the target speaker's speech segments in the audio with imbalanced speaker distribution, where the target speaker's voice predominates. 

Moreover, from the learning aspect, effectively modeling the correlation between facial appearance and voice characteristics is significant. It is crucial to not only generate accurate and robust single-modal (face or speech) embeddings, but also to establish cross-modal alignment. However, simple joint-modeling approach for these components can hardly avoid over-fitting, complicating the understanding of both single modality information and cross-modal correlations. This requires a smoother training strategy to ensure effectiveness and generalizability.

To address the challenges outlined above, we propose the ``Multi-stage Face-voice Association Learning with Keynote Speaker Diarization''~(MFV-KSD) framework. MFV-KSD integrates a keynote speaker diarization frontend and a three-stage training strategy. The keynote speaker diarization frontend is employed to filter out interfering speech inputs. It can solve the data problem by enabling more robust and accurate target speaker's embedding. The three-stage training strategy consists of intra-modal recognition training, inter-modal correlation training, and FAME adaptation. By leveraging the large-scale dataset, this strategy can balance and enhance the intra-modal learning ability and the inter-modal alignment ability. Our system achieves 19.9\% EER in FAME 2024 challenge.

\section{Related work}

\subsection{Biometric recognition}
As human faces and voices carry unique individual characteristics, various biometric authentication models have been developed for speech modality, face modality, or both of them~\cite{face_voice_biometric}.

\subsubsection{Face recognition}
Each human face reflects a unique combination of features such as eyes, nose, and mouth, which enable the ability to recognize each person based on their facial characteristics \cite{face_recognition}. Deep neural networks can extract facial traits into their representations. Among all solutions, convolutional neural networks (CNNs) are particularly popular in face recognition systems due to their robust capability to extract detailed local features, such as the ResNet model~\cite{deng2019retinaface}.

\subsubsection{Speaker recognition}
The human voice is shaped by unique individual physiological factors, including the shape of the vocal tract, accent, and rhythm~\cite{speech_survey}. Consequently, speaker recognition systems can be employed to identify individuals. For example, the x-vector model~\cite{xvector} is based on Time Delay Neural Network (TDNN) architecture. The 2D-CNN ResNet architecture was adapted for speaker recognition tasks, achieving better performance. Building on these advancements, the ECAPA-TDNN model~\cite{desplanques2020ecapa} was proposed, which improves upon the x-vector by replacing the original TDNN layers with multi-level 1D residual Res2Net modules to capture speaker embeddings effectively.
 
\subsubsection{Audio-visual identity recognition}
Audio-visual identity recognition systems enhance the effectiveness of biometric applications across various scenarios. Fusion techniques for audio and visual modalities in identity recognition are primarily performed at the feature level \cite{feat_fuse_1_attn, feat_fuse_2_fc, feat_fuse_3_fc, feat_fuse_4_attn, feat_fuse_5_attn,jiang2023target} and the score level \cite{score_fuse_1, score_fuse_2}. Feature-level fusion projects the combination of audio and visual representations into a shared embedding space, whereas score-level fusion aggregates the scores obtained from each modality into a final prediction score.

\vspace{-0.5em}
\subsection{Cross-modal speaker verification}
Cross-modal speaker verification aims to determine whether a static image and an audio signal represent the same individual. This task is motivated by the fact that a person's face and voice can be associated based on correlations between physical factors such as age, gender, and ethnicity \cite{cross_modal_based}. This audio-visual cross-modal speaker verification system can help retrieve face-less voice segments or silent video clips of the target identity.

A common approach in this verification task is to map inputs from different modalities into a shared space to achieve cross-modal retrieval \cite{feat_fuse_3_fc}. For instance, \cite{cross_modal_1} trained a cross-modal model with a supervised identity-matching task using a fusion embedding between facial and voice encoders. Similarly, \cite{cross_modal_is2020} proposed optimizing the cross-modal architecture to increase the cosine similarity between facial and voice embeddings within the same identity system. Without requiring identity labels, \cite{cross_modal_2} introduced a cross-modal self-supervised learning scheme using unlabeled talking face videos. Another crucial aspect of cross-modal identity verification is linguistic dependence, as demonstrated by \cite{cross_modal_language}, which showed that changing the language in audio input results in a performance drop.

\begin{figure*}[!ht]
    \centering
    \includegraphics[width=.95\linewidth]{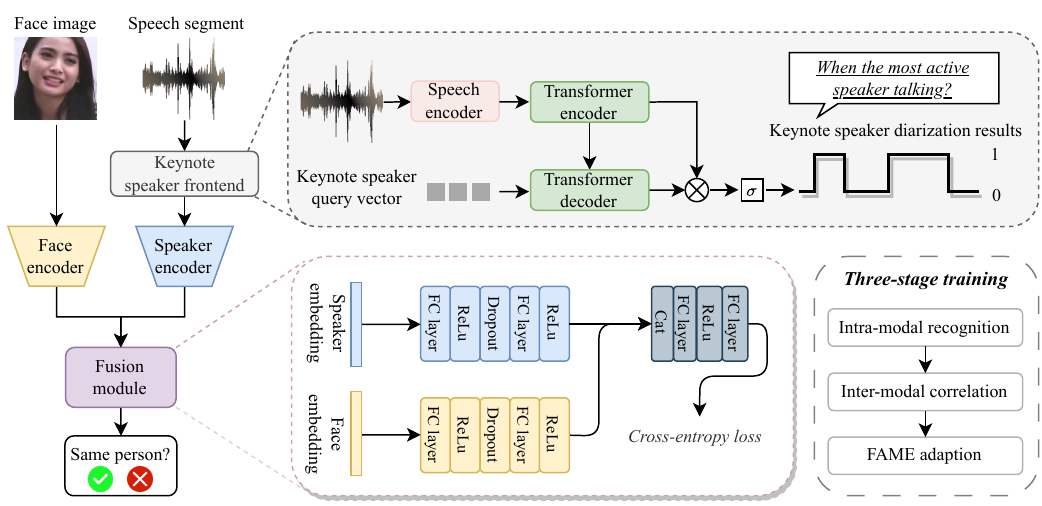}
    \caption{Multi-stage Face-voice Association Learning with Keynote Speaker Diarization (MFV-KSD) framework. It comprises a keynote speaker diarization frontend, face encoder, speaker encoder, and a fusion module. The training stage consists of three stages: intra-modal recognition training, inter-modal correlation training, and FAME adaption.}
    \label{fig:overview}
\end{figure*}

\vspace{-0.5em}
\section{Proposed Method: MFV-KSD}
In this section, we propose the Multi-stage Face-voice Association Learning with Keynote Speaker Diarization~(MFV-KSD) framework (Figure~\ref{fig:overview}) to solve the cross-modal speaker verification task. 

\vspace{-0.5em}
\subsection{Overall}
MFV-KSD consists of a keynote speaker frontend, face encoder, speaker encoder and a fusion module. The input is one face image and one speech segment, the output is the single score to represent if they come from the same person or not. As shown in the left side of Figure~\ref{fig:overview}, for the given audio and visual inputs, the keynote speaker diarization frontend is employed to detect the speech segments of the target speaker. Here, target speaker is the individual who speaks the most, i.e., keynote speaker, within the speech segment. Then the face image and speech segments of the target speaker are fed into the face encoder and speaker encoder, respectively, to obtain the face embedding and speaker embedding. This speaker encoder can extract the speaker embedding to capture the voice characteristics of keynote speaker. Similarly, the face encoder can obtain a face embedding to integrate the facial appearance.

Afterwards, we utilize a cross-modal fusion module to learn the correlation between audio and visual embeddings. The speaker and face embeddings are mapped into a shared space by several Fully Connected (FC) layers, with dropout layers incorporated to prevent over-fitting. We concatenate the multi-modal embeddings directly due to the efficiency consideration. 

% \vspace{-0.5em}
\subsection{Keynote speaker frontend}
Ideally, each audio utterance should contain speech from the target speaker only. However, in the real-world application, such as FAME 2024, the provided raw audios are conversations, which include speech segments from multiple speakers. That could affect the robustness of both training and prediction processes. Therefore, it is crucial to cleanse the samples and achieve high-quality inputs.

Traditional speaker diarization methods can hardly handle this condition due to the imbalanced distribution of speakers in each audio utterance. Motivated by our previous work in target speech diarization~\cite{jiang2024prompt}, we propose a keynote speaker diarization system by integrating a keynote speaker query vector, which is a learnable vector. We adopt the pre-trained keynote speaker diarization framework as frontend and use it to search the active segments of the target speaker. This frontend is pre-trained on the simulated conversation datasets and it is frozen in the MFV-KSD system .

As shown in the right-upper side of Figure~\ref{fig:overview}, this keynote speaker frontend takes speech segment and trained keynote speaker query vector as inputs and outputs a frame-level probability sequence, which represents the speaking activities of the keynote speaker in each frame. The transformer encoder inputs the speech feature extracted from speech encoder. The transformer decoder takes prompt vector and transformer encoder speech representation. Finally, we perform a dot product operation between the decoder output and the encoder output and apply a sigmoid operation to get the prediction sequence.

This keynote speaker frontend can filter the dataset effectively and minimize interference signals from the voices of non-relevant speakers, archiving more accurate model training and evaluation.

% \vspace{-0.5em}
\subsection{Three-stage training}
We propose a three-stage training pipeline for MFV-KSD to boost the system with the cross-modal learning ability. The initial intra-modal recognition training step aims to accurately capture biometric features from the single modality, i.e., facial or voice characteristics. Subsequently, inter-modal correlation training establishes the alignment between the facial appearance and voice characteristics from the large-scale dataset. Finally, FAME adaptation is applied to meet the in-domain evaluation. Note that our experiments have strictly met the unseen-language requirements.

\subsubsection{Intra-modal recognition training}
In the first training stage, our objective is to learn face representation and speaker representation for recognition purposes on the large-scale single-modal dataset.

For the speech modality input, we employ the ECAPA-TDNN~(Emphasized Channel Attention, Propagation and Aggregation in Time Delay Neural Network) model~\cite{desplanques2020ecapa} as the speaker encoder due to its reliable performance in identifying speakers. For the variable lengths of input utterances, the speaker encoder can learn the speaker embedding with the fixed dimension. In this training stage, the speaker embedding is fed into the classification layer to perform speaker recognition task with the AAM-softmax loss \cite{deng2019arcface}. The learnt embedding can represent the voice characteristic of the input speech.

For the face modality, we utilize the ResNet-18 model~\cite{deng2019arcface} as the face encoder to effectively extract detailed facial features and recognize the target speaker. The face encoder is also trained with a classification task in this stage. It can output face embeddings in MFV-KSD to capture robust facial appearance features. 

\vspace{-0.5em}
\subsubsection{Inter-modal correlation training}
Building on the pre-trained speaker and face encoders from the intra-modal recognition stage, we perform the inter-modal correlation training to establish a robust and general alignment between face and voice signals. This step links the facial features and voice embeddings by learning the large-scale samples in the wild. A fusion module is introduced to correlate the multi-modal biometric inputs. 

In this stage, each training data sample contains a facial image and corresponding reference speech from the same individual. They can offer biometric information from diverse perspectives. During training, the face component for half of the samples is replaced from the face of the different person to generate the negative pairs. Based on that, the pre-trained face encoder, speaker encoder, and fusion module can be trained with the cross-entropy loss for binary classification. This stage involves as many pairs as possible from the large audio-visual in-the-wild dataset that can guide the MFV-KSD to learn the general association.

\subsubsection{FAME adaptation}
Considering the domain mismatch between the pre-trained dataset and the FAME dataset, we further fine-tune the entire system on specific variants of the FAME dataset. This adaptation stage integrates our previously pre-trained framework into the evaluation domain, significantly enhancing the cross-modal speaker verification performance on the FAME dataset. A very small learning rate is used in this process.

\section{Experimental setup}
\subsection{FAME}
The 2024 Face-Voice Association in Multilingual Environments (FAME) challenge focuses on learning face-voice associations \cite{saeed2024face}. The goal of FAME is to verify whether a given sample, containing both a face and a voice, belongs to the same identity. This challenge specifically examines the impact of multiple languages. The trained model is evaluated on both seen and unseen languages to assess its performance in cross-language environments.

\vspace{-1.em}
\begin{table}[!hb]
\caption{Summary of datasets used for implementation}
\centering
\begin{tabular}{>{\centering\arraybackslash}p{2cm}>{\centering\arraybackslash}p{2cm}>{\centering\arraybackslash}p{2.5cm}}
\hline
Stage & Dataset & Modality \\ \hline
\multirow{2}{*}{\begin{tabular}[c]{@{}c@{}}Intra-modal\\ recognition\end{tabular}} & VoxCeleb2 & Speech \\ \cline{2-3} 
& Glint360K & Face                             \\ \hline
\multirow{2}{*}{\begin{tabular}[c]{@{}c@{}}Inter-modal\\ correlation\end{tabular}} & VoxCeleb1 & \multirow{2}{*}{Speech and face} \\
& MAV-Celeb &                                  \\ \hline
\begin{tabular}[c]{@{}c@{}}FAME\\ adaption\end{tabular}                      & MAV-Celeb & Speech and face                  \\ \hline
\end{tabular}
\label{tab:dataset}
\end{table}

\vspace{-1.em}

\subsection{Dataset}
As summarized in Table~\ref{tab:dataset}, we use large-scale VoxCeleb2 \cite{Voxceleb2} and Glint360K \cite{deng2019retinaface} datasets for intra-modal recognition training. Inter-modal correlation training applies the VoxCeleb1 \cite{Voxceleb} and MAV-Celeb datasets, both of which are audio-visual. MAV-Celeb is the dataset that used in FAME, it features 154 celebrities speaking three languages: English, Hindi, and Urdu. In the FAME adaptation stage, only MAV-Celeb dataset is used.

To meet the unseen language requirements, we cleanse VoxCeleb1 and VoxCeleb2 by VoxLingua107 \cite{valk2021voxlingua107} model, which provide language labels. For instance, for English-unseen, all three training stages are performed on VoxCeleb1, VoxCeleb2 or MAV-Celeb datasets without the English language to meet the requirements. To monitor the performance, we split 6 speakers from the training set to generate the validation dataset.

\section{Results}
\subsection{Performance in the FAME challenge}
In FAME 2024, our best results on the test set can be found in Table~\ref{tab:results}. MFV-KSD has outperformed the given baseline in FAME~\cite{saeed2024face} and achieved 19.9\% overall EER. Meanwhile, the heard results perform better than the unheard condition. That proves the importance of language in this task. Also, we found that the results for Urdu were much better than those for English.

\begin{table}[!h]
  \caption{Evaluation results (EER (\%)) on MAV-Celeb-V1 test}
  \centering
  \begin{tabular}{cccccc}
  \hline
    Method  & E-seen & E-unseen  & U-seen & U-unseen & Overall \\
    \hline
    Baseline~\cite{saeed2022fusion} & 29.3 & 37.9 & 25.8  & 40.4& 33.4 \\
    \textbf{MFV-KSD} & 21.8 & 27.3 & 14.7 & 15.8  & 19.9 \\
    \hline
  \end{tabular}
  \label{tab:results}
\end{table}

\vspace{-1.5em}
\begin{table}[!ht]
  \caption{Ablation study on the MAV-Celeb-V1 set for keynote speaker frontend (without inter-modal correlation training)}
  \centering
  \begin{tabular}{cccc}
  \hline
      Key-train & Key-test & English & Urdu\\
      \hline
      $\checkmark$ & $\checkmark$ & 28.0 & 15.1\\
      $\times$ & $\times$ & 29.9 &  17.0\\
      \hline
      $\checkmark$ & $\times$ & 28.8 & 14.7 \\
      $\times$ & $\checkmark$ & 30.3 & 17.4\\
    \hline
  \end{tabular}
  \label{tab:abl_key}
\end{table}

\vspace{-1.5em}
\begin{table}[!ht]
  \caption{Ablation study on the MAV-Celeb-V1 set for three-stage training (with keynote speaker frontend)}
  \centering
  \begin{tabular}{ccccc}
  \hline
      Intra-modal  & Inter-modal & FAME  & \multirow{2}{*}{English} & \multirow{2}{*}{Urdu} \\
      recognition & correlation & adaption & & \\
      \hline
      $\checkmark$ & $\checkmark$ & $\checkmark$ & 21.8 & 14.7\\
      $\checkmark$ & $\times$ & $\checkmark$     & 28.0 & 15.1\\
      $\checkmark$ & $\checkmark$ & $\times$ & 30.6 & 16.1 \\
    \hline
  \end{tabular}
  \label{tab:abl_pre}
\end{table}

\subsection{Ablation study}
For post-evaluation, we conduct the ablation study on the MAV-Celeb-V1 test set (seen condition only). 

\subsubsection{Keynote speaker frontend}
In Table~\ref{tab:abl_key}. `Key-train' and `Key-test' denote if we use this Keynote speaker frontend during training and testing, respectively. The results verify that this frontend can enhance the dataset and lead to clear improvement when it performs on the training set and test set together.

\subsubsection{Multi-stage training}
Then we study the effectiveness of our proposed three-stage training strategy in Table~\ref{tab:abl_pre}. Intra-modal correlation stage involves the information from the large-scale datasets to boost the generality of the system. FAME adaptation stage leads the model to understand the in-domain data further. Both of them are important for the entire framework.

\vspace{-1.5em}
\begin{table}[!ht]
  \caption{Ablation study on the MAV-Celeb-V1 set for model architecture (without inter-modal correlation training)}
  \vspace{-1em}
  \centering
  \begin{tabular}{cccc}
  \hline
      Speaker encoder & Face encoder & English & Urdu\\
      \hline
      ECAPA-TDNN & ResNet18 &  28.0 & 15.1\\
      ECAPA-TDNN & ResNet34 &  29.1 & 17.0\\
      ECAPA-TDNN & ResNet50 &  27.8 & 16.4\\
      \hline
      ECAPA-TDNN & ResNet18 &  28.0 & 15.1\\
      CAM++      & ResNet18 &  35.5 & 19.6\\
      ResNet34   & ResNet18 &  34.0 & 19.2\\
    \hline
  \end{tabular}
  \label{tab:abl_system}
\end{table}

\begin{table}[!ht]
  \caption{Ablation study on the learning rate issue (without inter-modal correlation training)}
  \vspace{-1em}
  \centering
  \begin{tabular}{cccc}
  \hline
      Learning-rate & 1e-5 & 1e-4 & 1e-3 \\
      \hline
      English & 28.0 & 29.8 & 33.8 \\
      Urdu   & 15.1 & 19.5 & 24.0 \\
    \hline
  \end{tabular}
  \label{tab:abl_lr}
\end{table}
\vspace{-2em}

\subsubsection{Model architecture}
We further study the different encoder architectures in Table~\ref{tab:abl_system}. For the face modality, a larger face encoder cannot boost the system performance in our experiments. For the speech modality, ECAPA-TDNN performs better than CAM++ \cite{wang2023cam++} and ResNet34 \cite{resnet_34}.

\subsubsection{Learning rate}
At last, we explore the importance of a small learning rate in Table~\ref{tab:abl_lr}. Since multi-stage training makes the system learn valuable knowledge, a very small learning rate can significantly boost the system.

\subsection{Quality study}

We further perform the quality study on our generated validation set to understand how MFV-KSD works in this task.

\subsubsection{Confusion matrix}
\begin{figure}[!ht]
    \centering
    \includegraphics[width=.85\linewidth]{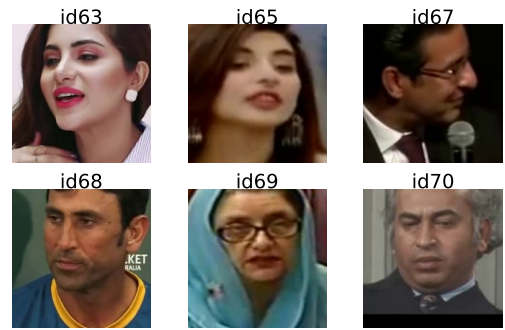}
    \vspace{-2em}
    \caption{Confusion matrix in the generated validation set.}
    \label{visual}
    \vspace{-1.em}
\end{figure}

First, we make the confusion matrix on the validation set, which contains 6 people in Figure~\ref{visual}. We find that MFV-KSD performs well for the cross-gender condition. However, the same gender people with similar age and nationality are very hard to distinguish (id63 vs id65). 

\begin{table}[!hb]
  \caption{Quality study of MFV-KSD for the gender effect}
  \vspace{-1em}
  \centering
  \begin{tabular}{cccc}
  \hline
      Validation set & Normal & Easy & Hard\\
      \hline
      English & 23.9 & 5.4 & 42.5\\
      Urdu  & 24.3 & 3.3 & 38.2\\
    \hline
  \end{tabular}
  \label{tab:vis}
\end{table}
\vspace{-1em}

\subsubsection{Gender effect}
\begin{figure}[!ht]
    \centering
\includegraphics[width=.8\linewidth]{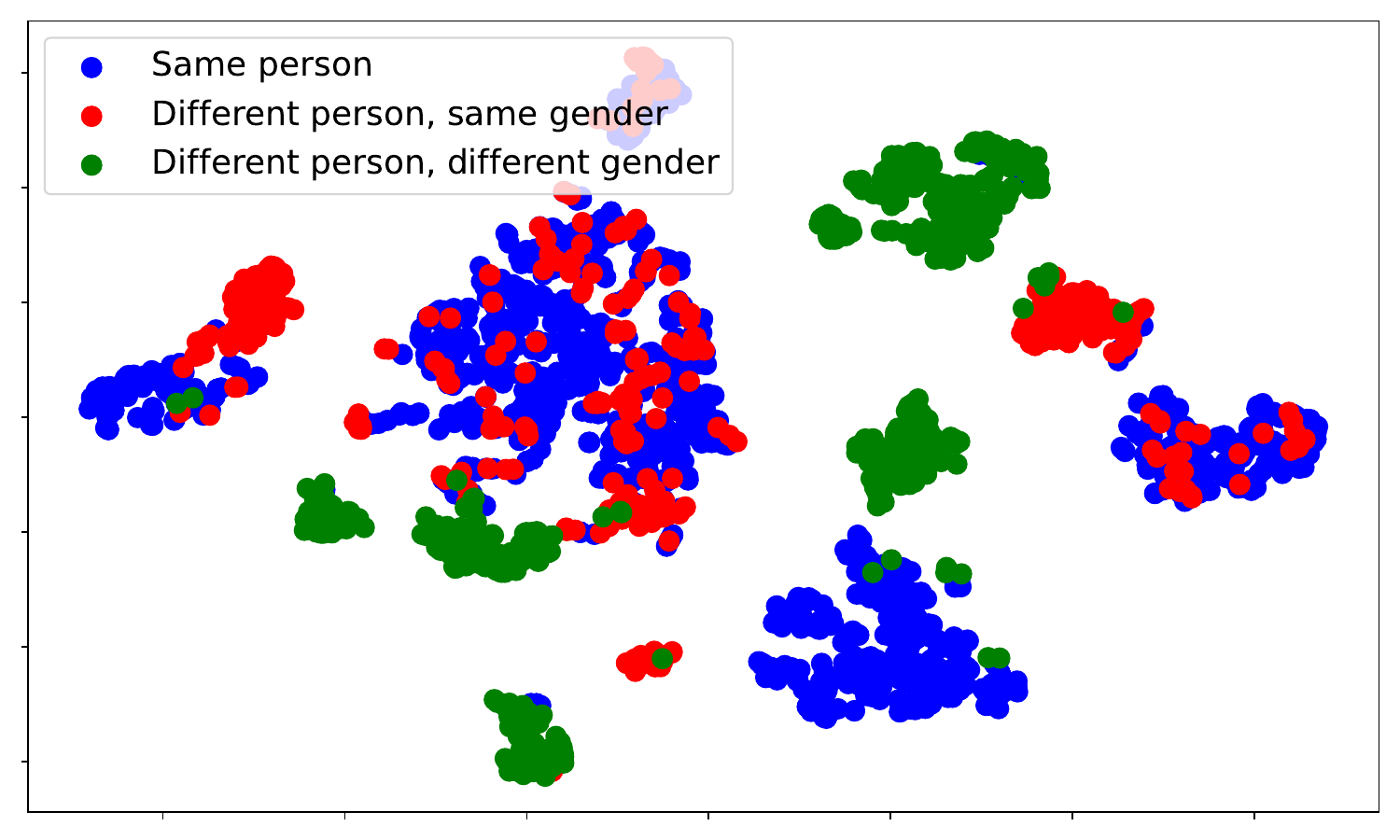}
    \vspace{-1.em}
    \caption{T-SNE visualization to study the gender effect.}
    \label{t-sne}
\end{figure}

In Table~\ref{tab:vis}, we set three validation sets: `Normal' has no restriction, `Easy' and `Hard' means all negative pairs are cross-gender or same-gender, respectively. System can do very well for the easy conditions but hard to work for these hard conditions. Figure~\ref{t-sne} conducts the T-SNE visualization for the `Normal' validation set based on the extracted multi-modal feature, pairs with the same person (blue dots) have the clear boundary with pairs with different persons, different gender (green dots). However, pairs with different persons, the same gender (red dots) are mixed with these blue dots. It is still challenging to handle the same-gender condition.

\vspace{-2mm}
\section{Conclusion}
This paper proposes a robust cross-modal speaker verification system to handle real-world challenging conditions. In future work, a standard benchmark is required to build, which contains the large-scale, fixed training set, evaluation set with enough samples, different tracks based on gender, age and nationality, and diverse tasks (verification, matching and retrieval). That can assist the community in comparing the various methods fairly.

\vspace{-2mm}
\section{Acknowledgement}
The research is supported by: National Natural Science Foundation of China, Grant No. 62271432; Shenzhen Science and Technology Program ZDSYS20230626091302006; Shenzhen Science and Technology Research Fund, Fundamental Research Key Project Grant No. JCYJ20220818103001002; FD-fAbrICS, Joint Lab for FD-SOI Always-on Intelligent \& Connected Systems, Award I2001E0053. Any opinions expressed in this material do not reflect the views of the institutions supporting the joint lab.

\clearpage
\balance
\footnotesize
\bibliographystyle{ACM-Reference-Format}
\bibliography{sample-base}
\end{document}